\documentclass[twocolumn,english,reprint, longbibliography, superscriptaddress, breaklinks=true, showkeys, showpacs=false, nofootinbib]{revtex4}
\usepackage[T1]{fontenc}
\usepackage[latin9]{inputenc}
\usepackage{color}
\usepackage{babel}
\usepackage{amsmath}
\usepackage{amssymb}
\usepackage{subfigure}
\usepackage{graphicx}
\usepackage{physics} 
\usepackage{hyperref}
\hypersetup{
    colorlinks=true,
    linkcolor=blue,
    filecolor=blue,      
    urlcolor=blue,
    }
\makeatletter
\@ifundefined{textcolor}{}
{%
 \definecolor{BLACK}{gray}{0}
 \definecolor{WHITE}{gray}{1}
 \definecolor{RED}{rgb}{1,0,0}
 \definecolor{GREEN}{rgb}{0,1,0}
 \definecolor{BLUE}{rgb}{0,0,1}
 \definecolor{CYAN}{cmyk}{1,0,0,0}
 \definecolor{MAGENTA}{cmyk}{0,1,0,0}
 \definecolor{YELLOW}{cmyk}{0,0,1,0}
}
\pdfoutput=1
\hypersetup{colorlinks=true,citecolor=blue,linkcolor=cyan,urlcolor=blue,filecolor= green, breaklinks=true}
\usepackage{url}
\usepackage{breakurl}
\makeatother
\usepackage{qcircuit}

\begin{document}

\title{Local predictability and coherence versus distributed entanglement \\ in entanglement swapping from partially entangled pure states}

\author{Jonas Maziero}
\email{jonas.maziero@ufsm.br}
\address{Departament of Physics, Center for Natural and Exact Sciences, Federal University of Santa Maria, Roraima Avenue 1000, Santa Maria, Rio Grande do Sul, 97105-900, Brazil}

\author{Marcos L. W. Basso}
\email{marcoslwbasso@hotmail.com}
\address{Center for Natural and Human Sciences, Federal University of ABC, Avenue of the States, Santo Andr\'e, S\~ao Paulo, 09210-580, Brazil}

\author{Lucas C. C\'eleri}
\email{lucas@qpequi.com}
\address{QPequi Group, Institute of Physics, Federal University of Goi\'as, Goi\^ania, Goi\'as, 74.690-900, Brazil}

\begin{abstract}
Complete complementarity relations, as e.g. $P(\rho_A)^2 + C(\rho_A)^2 + E(|\Psi\rangle_{AB})^2=1$, constrain the local predictability and local coherence and the entanglement of bipartite pure states. For pairs of qubits prepared initially in a particular class of partially entangled pure states with null local coherence, these relations were used in Ref. [Phys. Lett. A, 451, 128414 (2022)] to provide an operational connection between local predictability of the pre-measurement states with the probability of the maximally entangled components of the states after the Bell-basis measurement of the entanglement swapping protocol (ESP). In this article, we extend this result for general pure initial states. We obtain a general relation between pre-measurement local predictability and coherence and the distributed entanglement in the ESP. We use IBM's quantum computers to verify experimentally some instances of these general theoretical results. 
\end{abstract}


\maketitle
 
\section{Introduction} 

Distribution of quantum entanglement is an elementary operation for implementing quantum communication schemes. Entanglement swapping, proposed by Zukowski et al.~\cite{Anton}, is an elegant protocol to implement such operation. The original protocol starts with Alice and Charlie sharing a pair of maximally entangled qubits with Charlie also sharing another pair of maximally entangled qubits with a third partner, named Bob. Then, Charlie makes a Bell-basis measurement (BBM) on one qubit from each pair, which enables Alice and Bob to share a maximally entangled two-qubit state even though they possess qubits that possibly have never interacted before \cite{Zeilinger}. Bose et al. \cite{Knight} took a step forward and generalized entanglement swapping to a multi-qubit system while the authors in Ref.~\cite{Bouda} generalized the protocol to multi-qudit systems. On the other hand, the entanglement swapping from initially partially entangled states was first studied in Ref. \cite{Bose}. More recently, in Ref. \cite{pla_P_E}, two of us considered a more general class of initially partially entangled states and showed that there is a diversity of partially entangled states as well as maximally entangled states that can be obtained after a BBM. For the partially entangled states obtained after a BBM, it is possible in principle to purify them into a smaller number of maximally entangled pairs of quantum systems (quantons) \cite{Bennett}.  Beyond that, we studied the entanglement swapping for partially entangled states from a complementarity-entanglement relation perspective and showed that, in the cases where the entanglement increases after the BBM, the predictability is consumed when compared to the initially prepared state and we related the predictability of the pre-measurement one-qubit density matrix  to the probability of the partially entangled component of the post-measurement state. With this, it was then shown that predictability can be given not only a mathematical resource theory \cite{P_res}, but also an operational interpretation in a quantum communication protocol \cite{pla_P_E}. 

Complementarity-entanglement relations are also known as complete complementarity relations (CCR) and emerge in the context of the quantification of Bohr's complementarity principle~\cite{Bohr}, where the first quantitative version of the wave-particle duality was explored by Wootters and Zurek~\cite{Wootters} in the light of quantum information theory. Later, Greenberger and Yasin~\cite{Yasin} obtained an inequality, known as a complementarity relation, expressed in terms of a priori path-information, named predictability, and the interferometric visibility. Ever since, many approaches have been taken for quantifying the wave-particle perspective of a quanton in terms of complementarity relations~\cite{Engle, Ribeiro, Bera, Bagan, Coles, Hillery, Maziero, Leopoldo, Durr,Englert_}. For instance, it was realized that the quantum coherence~\cite{Baumgratz} is the natural generalization for the visibility measure for $d$-dimensional quantons (qudits)~\cite{Bera, Bagan, Mishra}. As well, it was uncovered by Jakob and Bergou~\cite{Janos} that entanglement has a important role in the wave-particle complementarity relation. These authors suggested the first triality relation, also known as CCR, involving predictability, quantum coherence and entanglement for a pair of qubits $AB$ in a bipartite pure state
\begin{align}
    P_A^2 + C_A^2 + E^2_{\textrm{conc}} = 1, \label{eq:ccr} 
\end{align}
where 
\begin{align}
& P_A:=P(\rho_A) := \abs{\Tr (\rho_A \sigma_z)}, \\ & C_A:=C(\rho_A) := 2 \abs{\Tr (\rho_A \sigma_+)},
\end{align}
with $\sigma_+ = \sigma_x + i \sigma_y$, where $\sigma_x,\sigma_y,\sigma_z$ are the Pauli's matrices, and $E_{\textrm{conc}} := E_{conc}(\ket{\Psi}_{AB})$ is the well known concurrence measure of entanglement~\cite{Willian} for a bipartite quantum system in the state $\ket{\Psi}_{AB}$ with $\rho_A = \Tr_B \ket{\Psi}_{AB}\bra{\Psi}$, where $Tr_B$ is the partial trace operation \cite{ptr}.
Besides, the measure of quantum coherence defined in Ref. \cite{Janos} is equivalent to the $l_1$-norm of quantum coherence~\cite{Baumgratz} for a qubit. It is worth mentioning that the CCRs were already generalized for multipartite pure qudits \cite{Marcos} as well as for bipartite mixed states~\cite{Jonas}, and many interesting properties of CCRs has been explored recently~\cite{BassoJPA, Qureshi, Wayhs, CCRin, CCRcst, CCRneu}. For instance, in Ref. \cite{BassoJPA} it was demonstrated that for each predictability and quantum coherence measures in a complementarity inequality satisfying the criteria established in Refs. \cite{Durr,Englert_}, it is possible to defined an entanglement monotone for bipartite pure states that completes the relation, turning the inequality into an equality. As well, the predictability measure together with the entanglement monotone can be combined to measure the path distinguishability in an interferometer \cite{Qureshi, Wayhs}, as was first discussed by Englert \cite{Engle}. Finally, such CCRs have been shown to have important invariance and dynamical properties in relativistic settings \cite{CCRin, CCRcst, CCRneu}.

In this article, in Sec. \ref{sec:gen}, we extend the results obtained in Ref. \cite{pla_P_E} by considering a general bipartite pure state for two qubits and establishing the relation between $P$, $C$ and the distributed entanglement in the entanglement swapping protocol by given a general formula for the post-measurement entanglement in terms of the local predictability and quantum coherence of the pre-measurement states. We notice that the averaged distributed entanglement is the  product of the initial entanglements, that, because of the restriction imposed by the CCRs, are determined by the corresponding local coherence and predictability. Besides, we look at some particular examples and learn that, in the cases where the post-measurement state is maximally entangled, the pre-measurement local coherence and local predictability are consumed, being transformed into one ebit of post-measurement entanglement. In Sec. \ref{sec:exp}, we verify experimentally some instances of these general results using IBM's quantum computers \cite{ibmq}.
Finally, in Sec. \ref{sec:conc}, we give our final remarks.

\section{Entanglement swap from partially entangled pure states}
\label{sec:gen}

Let us consider two pairs of qubits prepared in general pure states:
\begin{align}
|\xi\rangle_{AC} &= c_{00}|00\rangle +  c_{01}|01\rangle + c_{10}|10\rangle + c_{11}|11\rangle, \nonumber \\
|\eta\rangle_{C'B} &= d_{00}|00\rangle +  d_{01}|01\rangle + d_{10}|10\rangle + d_{11}|11\rangle.
\end{align}
The corresponding global state of the four qubits can be rewritten as follows
\begin{align}
& |\xi\rangle_{AC}|\eta\rangle_{C'B} \\
&=  ||\phi_+||\cdot|\Phi_+\rangle_{CC'}|\hat{\phi}_+\rangle_{AB} + ||\phi_-||\cdot|\Phi_-\rangle_{CC'}|\hat{\phi}_-\rangle_{AB} \nonumber \\
& + ||\psi_+||\cdot|\Psi_+\rangle_{CC'}|\hat{\psi}_+\rangle_{AB} + ||\psi_-||\cdot|\Psi_-\rangle_{CC'}|\hat{\psi}_-\rangle_{AB}, \nonumber
\end{align}
where we defined
\begin{align}
& |\hat{\phi}_+\rangle_{AB} := \frac{|\phi_+\rangle_{AB}}{||\phi_+||} :=  \frac{|\phi_+\rangle_{AB}}{\sqrt{Pr(\Phi_+)}}, \\
& |\hat{\phi}_-\rangle_{AB} := \frac{|\phi_-\rangle_{AB}}{||\phi_-||} := \frac{|\phi_-\rangle_{AB}}{\sqrt{Pr(\Phi_-)}}, \\
& |\hat{\psi}_+\rangle_{AB} := \frac{|\psi_+\rangle_{AB}}{||\psi_+||} := \frac{|\psi_+\rangle_{AB}}{\sqrt{Pr(\Psi_+)}}, \\
& |\hat{\psi}_-\rangle_{AB} := \frac{|\psi_-\rangle_{AB}}{||\psi_-||} := \frac{|\psi_-\rangle_{AB}}{\sqrt{Pr(\Psi_-)}},
\end{align}
with
\begin{align}
& 2^{1/2}|\phi_+\rangle_{AB} \\
& := (c_{00}d_{00}+c_{01}d_{10})|00\rangle + (c_{00}d_{01}+c_{01}d_{11})|01\rangle \nonumber \\
& + (c_{10}d_{00}+c_{11}d_{10})|10\rangle + (c_{10}d_{01}+c_{11}d_{11})|11\rangle, \nonumber\\
& 2^{1/2}|\phi_-\rangle_{AB} \\
& := (c_{00}d_{00}-c_{01}d_{10})|00\rangle + (c_{00}d_{01}-c_{01}d_{11})|01\rangle \nonumber\\
& + (c_{10}d_{00}-c_{11}d_{10})|10\rangle + (c_{10}d_{01}-c_{11}d_{11})|11\rangle, \nonumber \\
& 2^{1/2}|\psi_+\rangle_{AB} \\
& := (c_{00}d_{10}+c_{01}d_{00})|00\rangle + (c_{00}d_{11}+c_{01}d_{01})|01\rangle \nonumber \\
& + (c_{10}d_{10}+c_{11}d_{00})|10\rangle + (c_{10}d_{11}+c_{11}d_{01})|11\rangle, \nonumber \\
& 2^{1/2}|\psi_-\rangle_{AB} \\
& := (c_{00}d_{10}-c_{01}d_{00})|00\rangle + (c_{00}d_{11}-c_{01}d_{01})|01\rangle \nonumber\\
& + (c_{10}d_{10}-c_{11}d_{00})|10\rangle + (c_{10}d_{11}-c_{11}d_{01})|11\rangle \nonumber
\end{align}
and
\begin{align}
& 2Pr(\Phi_+) = 2||\phi_+||^{2} \label{bbmP1} \\
& = |c_{00}d_{00}+c_{01}d_{10}|^{2} + |c_{00}d_{01}+c_{01}d_{11}|^{2} \nonumber \\
& + |c_{10}d_{00}+c_{11}d_{10}|^{2} + |c_{10}d_{01}+c_{11}d_{11}|^{2}, \nonumber \\
& 2Pr(\Phi_-) = 2||\phi_-||^{2} \\
& = |c_{00}d_{00}-c_{01}d_{10}|^{2} + |c_{00}d_{01}-c_{01}d_{11}|^{2} \nonumber \\
& + |c_{10}d_{00}-c_{11}d_{10}|^{2} + |c_{10}d_{01}-c_{11}d_{11}|^{2}, \nonumber \\
& 2Pr(\Psi_+) = 2||\psi_+||^{2} \\
& = |c_{00}d_{10}+c_{01}d_{00}|^{2} + |c_{00}d_{11}+c_{01}d_{01}|^{2} \nonumber \\
& + |c_{10}d_{10}+c_{11}d_{00}|^{2} + |c_{10}d_{11}+c_{11}d_{01}|^{2}, \nonumber \\
& 2Pr(\Psi_-) = 2||\psi_-||^{2} \label{bbmP4} \\
& = |c_{00}d_{10}-c_{01}d_{00}|^{2} + |c_{00}d_{11}-c_{01}d_{01}|^{2} \nonumber \\
& + |c_{10}d_{10}-c_{11}d_{00}|^{2} + |c_{10}d_{11}-c_{11}d_{01}|^{2}. \nonumber 
\end{align}

Using the fact that for a two-qubit pure state $|\Phi\rangle=\alpha_{00}|00\rangle+\alpha_{01}|01\rangle+\alpha_{10}|10\rangle+\alpha_{11}|11\rangle$ the entanglement concurrence \cite{concurrence} is given by $E_{con}(\Phi)=2|\alpha_{00}\alpha_{11}-\alpha_{01}\alpha_{10}|$, we can write the entanglement of the post-measurement states of the qubits $A$ and $B$ in the entanglement swapping protocol as
\begin{align}
E_{con}(\hat{\phi}_+) & = \frac{E_{con}(\xi_{AC})E_{con}(\eta_{C'B})}{4Pr(\Phi_+)}, \\
E_{con}(\hat{\phi}_-) & = \frac{E_{con}(\xi_{AC})E_{con}(\eta_{C'B})}{4Pr(\Phi_-)}, \\
E_{con}(\hat{\psi}_+) & = \frac{E_{con}(\xi_{AC})E_{con}(\eta_{C'B})}{4Pr(\Psi_+)}, \\
E_{con}(\hat{\psi}_-) & = \frac{E_{con}(\xi_{AC})E_{con}(\eta_{C'B})}{4Pr(\Psi_-)}.
\end{align}
So, the averaged entanglement of the qubits $A$ and $B$ after a non-selective Bell-basis measurement on the qubits $C$ and $C'$ is
\begin{align}
\langle E_{conc}\rangle =&  E_{con}(\hat{\phi}_+)Pr(\hat{\phi}_+) + E_{con}(\hat{\phi}_-)Pr(\hat{\phi}_-) \\
& + E_{con}(\hat{\psi}_+)Pr(\hat{\psi}_+) + E_{con}(\hat{\psi}_-)Pr(\hat{\psi}_-) \\
= & E_{con}(\xi_{AC})E_{con}(\eta_{C'B}) \\
= & \sqrt{1-P_C^2-C_C^2}\sqrt{1-P_{C'}^2-C_{C'}^2},
\end{align}
where we used $P_C:=P(\xi_C)$, $C_C:=C_{l_1}(\xi_C)$, $P_{C'}:=P(\xi_{C'})$, $C_{C'}:=C_{l_1}(\xi_{C'})$. So, we see that the averaged distributed entanglement is simple the product of the initial entanglements, that, because of the restriction imposed by the CCRs, are determined by the corresponding local coherence and predictability. In the sequence, we shall regard post-selection in the BBM and we will write the entanglement of each post-measurement state of the qubits $A$ and $B$ in terms of the local predictabilities and coherences.

After some algebra, we can express the BBM probabilities of Eqs. (\ref{bbmP1})-(\ref{bbmP4}) as
\begin{align}
4Pr(\Phi_+) & = 2\big(\xi^C_{00}\eta^{C'}_{00} + \xi^C_{11}\eta^{C'}_{11}\big) + 4\Re\big(\xi_{01}^C\eta_{01}^{C'}\big) \\
& = 1+(\xi_{00}^C-\xi_{11}^C)(\eta_{00}^{C'}-\eta_{11}^{C'}) + 4\Re\big(\xi_{01}^C\eta_{01}^{C'}\big), \nonumber \\
4Pr(\Phi_-) & = 2\big(\xi^C_{00}\eta^{C'}_{00} + \xi^C_{11}\eta^{C'}_{11}\big) - 4\Re\big(\xi_{01}^C\eta_{01}^{C'}\big) \\
& = 1+(\xi_{00}^C-\xi_{11}^C)(\eta_{00}^{C'}-\eta_{11}^{C'}) - 4\Re\big(\xi_{01}^C\eta_{01}^{C'}\big), \nonumber \\
4Pr(\Psi_+) & = 2\big(\xi_{00}^C\eta_{11}^{C'} + \xi_{11}^C\eta_{00}^{C'}\big)  + 4\Re\big(\xi_{01}^C\eta_{10}^{C'}\big) \\
& = 1-(\xi_{00}^C-\xi_{11}^C)(\eta_{00}^{C'}-\eta_{11}^{C'}) + 4\Re\big(\xi_{01}^C\eta_{10}^{C'}\big), \nonumber \\
4Pr(\Psi_-) & = 2\big(\xi_{00}^C\eta_{11}^{C'} + \xi_{11}^C\eta_{00}^{C'}\big)  - 4\Re\big(\xi_{01}^C\eta_{10}^{C'}\big) \\
& = 1-(\xi_{00}^C-\xi_{11}^C)(\eta_{00}^{C'}-\eta_{11}^{C'}) - 4\Re\big(\xi_{01}^C\eta_{10}^{C'}\big). \nonumber
\end{align}
The terms 
\begin{align}
\xi_{00}^C-\xi_{11}^C & = sgn(\xi_{00}^C-\xi_{11}^C)|\xi_{00}^C-\xi_{11}^C| \nonumber \\
& = sgn(\xi_{00}^C-\xi_{11}^C)P(\xi_C), \\
\eta_{00}^{C'}-\eta_{11}^{C'} & = sgn(\eta_{00}^{C'}-\eta_{11}^{C'})|\eta_{00}^{C'}-\eta_{11}^{C'}| \nonumber \\
& = sgn(\eta_{00}^{C'}-\eta_{11}^{C'})P(\eta_{C'})
\end{align}
are directly related to the local predictabilities. Above we used
$sgn(x)=\begin{cases}-1,\ x<0 \\ 0,\ x=0 \\ +1,\ x>0\end{cases}$.
We shall see that the terms $\Re\big(\xi_{01}^C\eta_{01}^{C'}\big)$ and $\Re\big(\xi_{01}^C\eta_{10}^{C'}\big)$ are related with the local coherences as follows. First we notice that any one-qubit state with Bloch representation \cite{Nielsen,Wilde} $\rho = \frac{1}{2}\begin{bmatrix} 1+r_z & r_x-ir_y \\ r_x+ir_y & 1-r_z \end{bmatrix}$, where $r_x=r\sin\theta\cos\phi$, $r_y=r\sin\theta\sin\phi$, and $r_z=r\cos\theta$, 
can be transformed, via the application of a rotation $R_z(-\phi)=e^{i\phi\sigma_z/2}$, into a state
$\rho' = \frac{1}{2}\begin{bmatrix} 1+r_z' & r_x' \\ r_x' & 1-r_z' \end{bmatrix}$ without changing the populations and coherence, i.e., with
with $r_z = r_z'$ and $r_x^2 + r_y^2 = (r_x')^2=r^2\sin^2\theta$.
So, if we define
\begin{align}
\tilde{\xi}_{AC} & = \big(\mathbb{I}_{A}\otimes R_Z(-\phi_C)\big)\xi_{AC}\big(\mathbb{I}_{A}\otimes R_Z(-\phi_C)^{\dagger}\big), \\
\tilde{\eta}_{C'B} & = \big(R_Z(-\phi_C')\otimes\mathbb{I}_{B}\big)\eta_{C'B}\big(R_Z(-\phi_C')^{\dagger}\otimes\mathbb{I}_{B}\big),
\end{align}
with $\phi_C$ and $\phi_{C'}$ being the azymuthal angle of the Bloch representation of the local state $\xi_C$ and $\eta_{C'}$, respectively, we get 
$\Im\big(\tilde{\eta}_{01}^{C'}\big)=\Im\big(\tilde{\eta}_{10}^{C'}\big)=0$. Therefore 
\begin{align}
4\Re\big(\tilde{\xi}_{01}^C\tilde{\eta}_{01}^{C'}\big) & = 4\Re\big(\tilde{\xi}_{01}^C\big)\Re\big(\tilde{\eta}_{01}^{C'}\big) = 2|\tilde{\xi}_{01}^C|2|\tilde{\eta}_{01}^{C'}|  \\
& = C_{l_1}(\tilde{\xi}_C)C_{l_1}(\tilde{\eta}_{C'}) = C_{l_1}(\xi_C)C_{l_1}(\eta_{C'}). \nonumber
\end{align}
With this, we relate the local predictability and coherence of the state before the BBM with entanglement of the state after the BBM:
\begin{align}
E_{con}^\pm(\hat{\phi}_+) & = \frac{\sqrt{\big(1-P_C^2-C_C^2\big)\big(1-P_{C'}^2-C_{C'}^2\big)}}{1\pm P_C P_{C'} + C_C C_{C'}}, \label{E1} \\
E_{con}^\pm(\hat{\phi}_-) & = \frac{\sqrt{\big(1-P_C^2-C_C^2\big)\big(1-P_{C'}^2-C_{C'}^2\big)}}{1\pm P_C P_{C'} - C_C C_{C'}}, \label{E2} \\
E_{con}^\pm(\hat{\psi}_+) & = \frac{\sqrt{\big(1-P_C^2-C_C^2\big)\big(1-P_{C'}^2-C_{C'}^2\big)}}{1\mp P_C P_{C'} + C_C C_{C'}}, \label{E3} \\
E_{con}^\pm(\hat{\psi}_-) & = \frac{\sqrt{\big(1-P_C^2-C_C^2\big)\big(1-P_{C'}^2-C_{C'}^2\big)}}{1\mp P_C P_{C'} - C_C C_{C'}}. \label{E4}
\end{align}
Above $E_{conc}^{+}$ stands for the cases where $\xi_C$ and $\eta_{C'}$ are in the same hemisphere of the Bloch ball while $E_{conc}^{-}$ applies when the local states are in opposite hemispheres of the Bloch ball. With this we have given a general formula for the post-measurement entanglement in terms of the local predictability and coherence of the pre-measurement states. Below we will look at some particular examples.

For initially maximally entangled qubit pairs, we have that $P_C=P_{C'}=C_C=C_{C'}=0$. Therefore, from Eqs. (\ref{E1})-(\ref{E4}), we have $E_{con}^\pm(\hat{\phi}_{\pm})=E_{con}^\pm(\hat{\psi}_{\pm})=1$, as is well known from the original ESP. 

Next, let us assume that $P_C=P_{C'}=0$ and $C_C=C_{C'}$. In this case we have that $E_{con}^\pm(\hat{\phi}_+)=E_{con}^\pm(\hat{\psi}_+)=(1-C_C^2)/(1+C_C^2)\in[0,1]$, with these entanglements being equal to one for $C_C=0$ and equal to zero for $C_C=1$, and $E_{con}^\pm(\hat{\phi}_-)=E_{con}^\pm(\hat{\psi}_-)=1$. On the other hand, if $C_C=C_{C'}=0$ and $P_C=P_{C'}$ then we get $E_{con}^+(\hat{\phi}_+)=E_{con}^+(\hat{\phi}_-)=(1-P_C^2)/(1+P_C^2)$, $E_{con}^+(\hat{\psi}_+)=E_{con}^+(\hat{\psi}_-)=1$, 
$E_{con}^-(\hat{\phi}_+)=E_{con}^-(\hat{\phi}_-)=1$, and $E_{con}^-(\hat{\psi}_+)=E_{con}^-(\hat{\phi}_-)=(1-P_C^2)/(1+P_C^2)$. Therefore, in all these cases we see that two of the post-measurement states of qubits $A$ and $B$ are maximally entangled. For these post-measurement states, one can say that the local predictability or local coherence of the states before the BBM was consumed, resulting in two ebits of entanglement after the BBM.

One can also verify that for $C_C=C_{C'}$ and $P_C=P_{C'}$ it follows that $E_{con}^+(\hat{\phi}_+)=(1-P_C^2-C_C^2)/(1+P_C^2+C_C^2)$, $E_{con}^+(\hat{\phi}_-)=(1-P_C^2-C_C^2)/(1+P_C^2-C_C^2)$, $E_{con}^+(\hat{\psi}_+)=(1-P_C^2-C_C^2)/(1-P_C^2+C_C^2)$, $E_{con}^+(\hat{\phi}_+)=1$,
$E_{con}^-(\hat{\phi}_+)=(1-P_C^2-C_C^2)/(1-P_C^2+C_C^2)$, $E_{con}^+(\hat{\phi}_-)=1$, $E_{con}^+(\hat{\psi}_+)=(1-P_C^2-C_C^2)/(1+P_C^2+C_C^2)$, and $E_{con}^+(\hat{\phi}_+)=(1-P_C^2-C_C^2)/(1+P_C^2-C_C^2)$. We see then that in these cases one post-measurement state is maximally entangled. That is to say, for this post-measurement state both pre-measurement local coherence and local predictability are consumed being transformed into one ebit of post-measurement entanglement. 

Now, let us consider as an explicit example the initial states regarded in Ref. \cite{pla_P_E}, i.e., we set $c_{00}=\sqrt{p}$, $c_{11}=\sqrt{1-p}$, $d_{00}=\sqrt{q}$, $d_{11}=\sqrt{1-q}$, with $p,q\in[0,1]$, and $c_{01}=c_{10}=d_{01}=d_{10}=0$. In this case we get $P_C = |2p-1|$, $P_{C'} = |2q-1|$, $C_C=C_{C'}=0$, $E_{conc}(\xi_{AC}) = 2\sqrt{p(1-p)}$, and
$E_{conc}(\eta_{C'B}) = 2\sqrt{q(1-q)}$. Thus, we obtain
\begin{align}
E_{con}^\pm(\hat{\phi}_+) & = E_{con}^\pm(\hat{\phi}_-) = \frac{2\sqrt{p(1-p)q(1-q)}}{pq+(1-p)(1-q)}, \label{EC01} \\
E_{con}^\pm(\hat{\psi}_+) & = E_{con}^\pm(\hat{\psi}_-) = \frac{2\sqrt{p(1-p)q(1-q)}}{p(1-q)+(1-p)q}. \label{EC03}
\end{align}
These expressions coincide with the ones presented in Ref. \cite{pla_P_E} in both cases $(p>1/2,q<1/2)$ and $(p<1/2,q>1/2)$. So, all the conclusions drawn in that article are obtained as a particular case of the results reported in the present paper.

In order to exemplify the analog role of coherence,  let us regard another explicit example with the pre-measurement states being
\begin{align}
& |\xi\rangle_{AC}=\sqrt{p}|++\rangle_{AC}+\sqrt{1-p}|--\rangle_{AC}, \label{psi1} \\ 
& |\eta\rangle_{C'B}=\sqrt{q}|++\rangle_{C'B}+\sqrt{1-q}|--\rangle_{C'B}, \label{psi2}
\end{align}
with $|\pm\rangle=2^{-1/2}\big(|0\rangle\pm|1\rangle\big)$.
In this case we have $P_C = P_{C'} = 0$, $C_C = |2p-1|$, and $C_{C'} = |2q-1|$. As the states used in this example are obtained from the states used in the example presented in the last paragraph by applying local Hadamard gates, the entanglements are the same in the two examples. Thus, as was shown in Ref. \cite{pla_P_E}, the post-measurement entanglement of the qubits $A$ and $B$ is maximized with $p=q$ or $p=1-q$. For instance, if we set $p=1-q$ the global pre-measurement state is written as follows
\begin{align}
& |\xi\rangle_{CC'}\otimes|\eta\rangle_{AB} = \label{eq:psi_global} \\
& \sqrt{(1-q)q}|\Phi_{+}\rangle_{CC'}|\Phi_{+}\rangle_{AB} + \sqrt{(1-q)q}|\Psi_+\rangle_{CC'}|\Psi_+\rangle_{AB} \nonumber \\
& + |\Phi_-\rangle_{CC'}\frac{1}{\sqrt{2}}\big(\sqrt{(1-q)^{2}}|+-\rangle_{AB}+\sqrt{q^{2}}|-+\rangle_{AB}\big) \nonumber \\
& - |\Psi_{-}\rangle_{CC'}\frac{1}{\sqrt{2}}\big(\sqrt{(1-q)^{2}}|+-\rangle_{AB}-\sqrt{q^{2}}|-+\rangle_{AB}\big). \nonumber
\end{align}
We notice in this global state that if post-selection is made on the BBM results, we can obtain two pairs of maximally entangled states. Because of the complete complementarity relation,
\begin{equation}
C_{l_1}(\xi_C)^2+E_{conc}(\xi_{AC})^2=1\xrightarrow[]{BBM} E_{conc}(AB)^2=1,
\end{equation}
we see that the pre-measurement local coherence is consumed to give two ebits of entanglement after the BBM. It is interesting to notice that we can end up with two pairs of maximally entangled states even for initial pairs of qubits sharing very little entanglement, but with high local coherence. Notwithstanding, the greater the initial local coherence the lesser the probability of the maximally entangled components of the global state. In this specific example, we have that
\begin{equation}
    C_C^2 = 1-2\big(Pr(|\Phi_+\rangle_{CC'})+Pr(|\Psi_+\rangle_{CC'})\big).
\end{equation}

\section{Experiment}
\label{sec:exp}

\begin{figure}[b]
\centering
\includegraphics[width=0.48\textwidth]{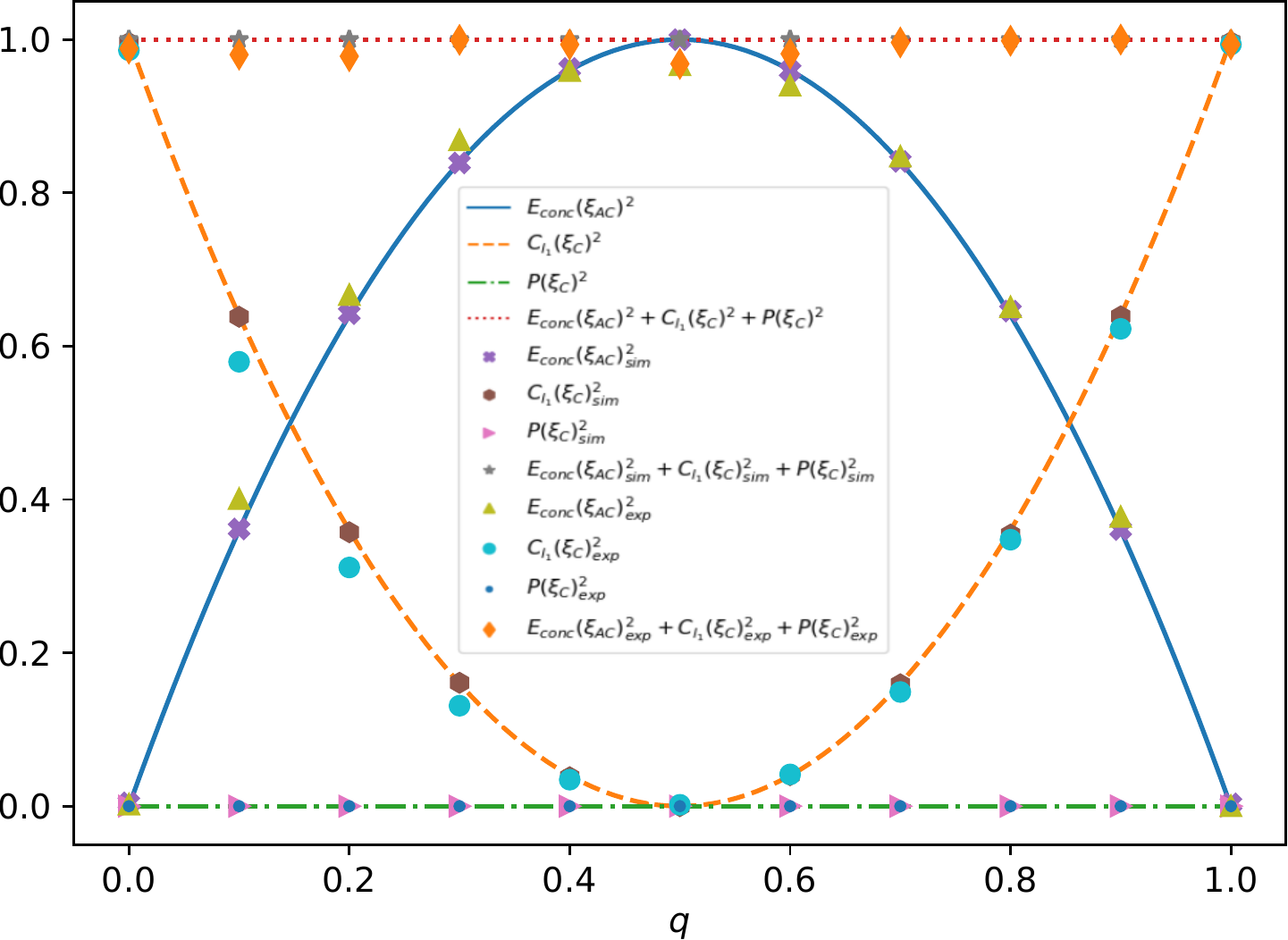}
\caption{Local coherence, predictability and entanglement of the qubit pairs states before the Bell basis measurement in the entanglement swapping protocol with generally partial entangled initial states. The lines are for the theoretical predictions and the points stand for the simulation and experimental results.}
\label{fig1}
\end{figure}

In this section, we use IBM's quantum computers to verify experimentally the last example discussed in the previous section. More specifically, we utilize the \textit{ibmq\underline\ manila} IBM quantum chip. We start by preparing the two pairs of initial states of Eqs. (\ref{psi1}) and (\ref{psi2}), with $p=1-q$, for $q\in[0,1]$. Using Qiskit's ready to use functions (see Ref. \cite{qiskit} and references therein), we make state tomography of these states. The theoretical, simulation and experimental values of the local coherences, predictabilities and the entanglements for these states are shown in Fig. \ref{fig1}. We applied measurement error mitigation \cite{qiskit} to all the experimental results we present in this article. This technique considerably improved our experimental results.

\begin{figure}[t]
\centering
\includegraphics[width=0.48\textwidth]{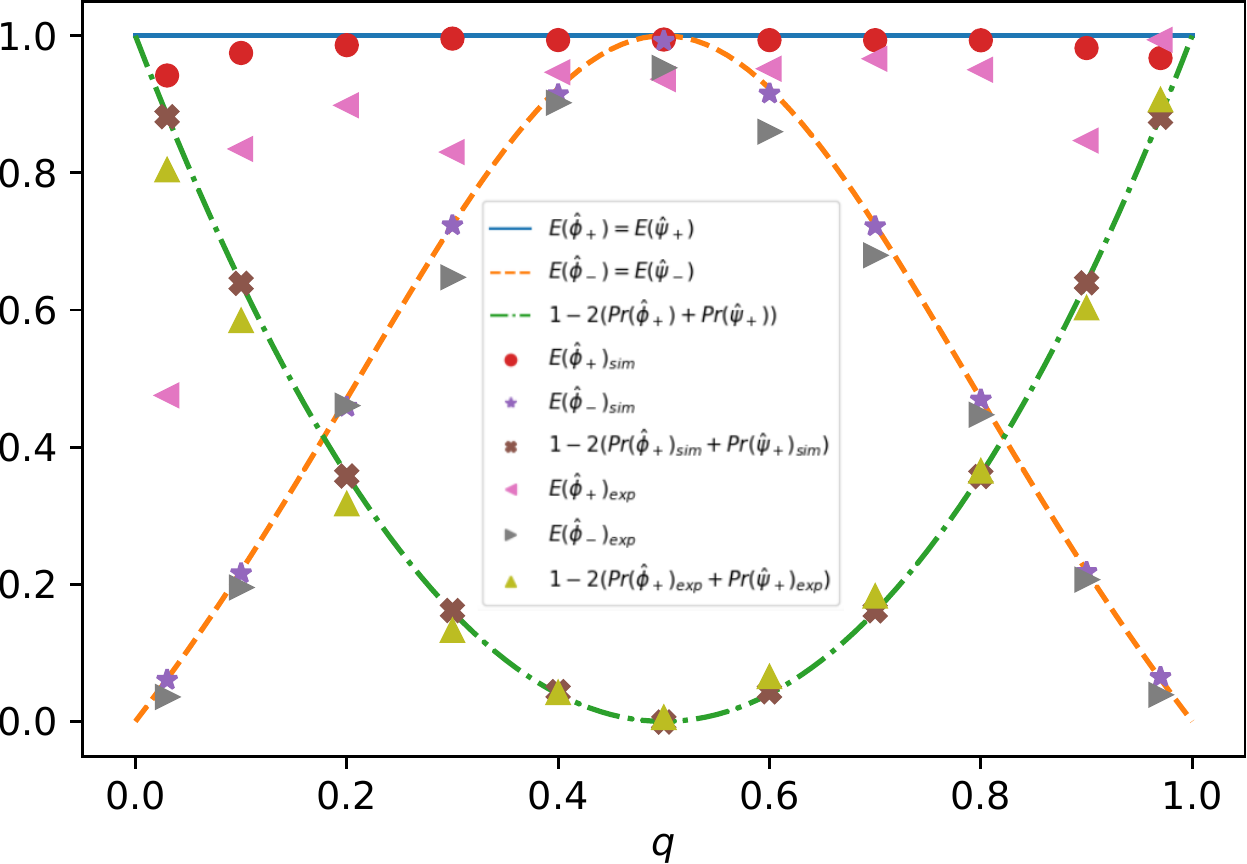}
\caption{Entanglement of the post-selected maximally and partially entangled components of the global state. The probabilities function $C_C^2 = 1-2\big(Pr(|\Phi_+\rangle_{AB})+Pr(|\Psi_+\rangle_{AB})\big)$ is shown too. The lines are for the theoretical predictions and the points stand for the simulation and experimental results.}
\label{fig2}
\end{figure}

Next, we prepare the states in Eqs. (\ref{psi1}) and (\ref{psi2}) again, for the same range of values of $q$, and we make a Bell basis measurement (BBM) on the qubits $C,C'$. In the sequence, with post-selection on the BBM results, we make state tomography of qubits $A,B$. With this we can compute the entanglement of each post-selected post-measurement state, verifying the transformation of local coherences into entanglement, and also verifying the relation of local coherence with the probability of the maximally entangled components of the global initial state. The theoretical, simulation, and experimental results are shown in Fig. \ref{fig2}.

In all cases the simulation values agree very well with the theoretical prediction. The experimental results also match fairly well with the theory. We observe that for the results in Fig. \ref{fig2}, the first and last values of $q$ were set to $0.03$ and $0.97$, respectively. This was done so in order to avoid numerical issues associated with the calculation of post-selected states appearing with small probability in Eq. (\ref{eq:psi_global}).

\section{Final remarks}
\label{sec:conc}

In this article, we investigated the entanglement swapping protocol (ESP) starting with pairs of qubits prepared in general, possibly partially entangled, pure states. We obtained general expressions relating the entanglement of the post-measurement states with the local coherence and predictability of the pre-measurement states. Via some examples, and using complete complementarity relations, we showed that, when at the end of the ESP one obtains two pairs of qubits in maximally entangled states, the pre-measurement local coherence and/or predictability is consumed and transformed into post-measurement entanglement. This can happen even for initial states with very little entanglement but large local coherence and/or predictability. However, the greater the initial local predictability and/or coherence, the lesser the probability of the maximally entangled component of the global pre-measurement state.

We used IBM's quantum computer as a quantum simulator and we tested experimentally some instances of our general theoretical results. The obtained simulation results agreed quite well with theory. Even with the high noise rate of current quantum computers, using measurement error mitigation, we obtained experimental results with fairly good agreement to the theoretical predictions.

Besides the contribution of our article from the fundamentals of quantum physics point of view, since it gives a complementarity-based operational connection of local coherence and/or predictability with the distributed entanglement in the ESP, we expect that it can also motivate practical developments related to entanglement distillation \cite{Bennett}
and to the usefulness of using partially entangled states for implementing quantum communication protocols under decoherence \cite{streltsov}.
 
\begin{acknowledgments}
This work was supported by the  S\~ao Paulo Research Foundation (FAPESP), Grant No.~2022/09496-8, by the National Institute for the Science and Technology of Quantum Information (INCT-IQ), process 465469/2014-0, and by the National Council for Scientific and Technological Development (CNPq), grants 309862/2021-3 and 304546/2019-4.
\end{acknowledgments}

\end{document}